\input harvmac
\baselineskip12pt
\noblackbox


\global\newcount\propno \global\propno=1
\def\prop#1{\xdef #1{\secsym\the\propno}\writedef{#1\leftbracket#1}%
{\bf\the\secno.}{\bf\the\propno.}
\global\advance\propno by1
\proplabeL#1}

\def\proplabeL#1{} 

\def\IF{{\bf F}}
\def\IP{{\bf P}}

\def\IC{{\bf C}}
\def\IQ{{\bf Q}}
\def\ZZ{{\bf Z}}
\def\pd{\partial}

\def\({ \left(  }
\def\){ \right) }

\def\tx{\theta_x}
\def\ty{\theta_y}
\def\tz{\theta_z}

\def\X#1#2(#3)#4#5{ {$X_{#1}^#2(#3)_{#5}^{#4}$} }

\def\pdp{{\pd \over \pd t_p}}
\def\cicy#1[#2|#3]#4{\left(\matrix{#2}\right|\!\!
                     \left|\matrix{#3}\right)^{{#4}}_{#1}}

%

\def\BCOVi{[1]}
\def\refBCOVi{ M. Bershadsky, S. Cecotti, H. Ooguri and C. Vafa, 
   {\it Holomorphic anomalies in topological field theories}, 
   (with an appendix by S.Katz) Nucl. Phys. B405(1993), 279-304.} 

\def\BCOVii{[2]}
\def\refBCOVii{ M. Bershadsky, S. Cecotti, H. Ooguri and C. Vafa, 
   {\it Kodaira-Spencer theory of gravity and exact results for quantum string
      amplitudes}, Commun. Math. Phys. 165(1994), 311-428.}

\def\Bat{[3]}
\def\refBat{
  V.V. Batyrev, {\it Dual polyhedra and mirror symmetry for Calabi-Yau 
  hypersurfaces in toric varieties}, J. Alg. Geom. 3(1994), 493-535.}

\def\BM{[4]}
\def\refBM{
    K. Behrend and Yu. Manin, 
    {\it Stacks of stable maps and 
    Gromov-Witten invariants}, Duke Math. J. 85(1996), 1-60. }

\def\BL{[5]}
\def\refBL{J. Bryan and N.C. Leoung, 
    {\it The enumerative geometry of $K3$ surfaces and modular forms}, 
    alg-geom/9711031.} 

\def\CdGP{[6]}
\def\refCdGP{ 
  P. Candelas, X.C. de la Ossa, P.S. Green, and L.Parkes, 
  {\it A pair of Calabi-Yau manifolds as an exactly soluble 
  superconformal theory}, Nucl.Phys. B356(1991), 21-74. }

\def\CN{[7]}
\def\refCN{ 
  J.H. Conway and S.P. Norton, 
  {\it Monstrous moonshine}, BLMS11 (1979), 308-339. }

\def\FP{[8]}
\def\refFP{
 C. Faber, R. Pandharipande, {\it Hodge integrals and Gromov-Witten theory}, 
 math/9810173.}

\def\G{[9]}
\def\refG{
  L. G\"ottsche, {\it The Botti numbers of the Hilbert scheme of points on 
  a smooth projective surface}, Math. Ann. 286 (1990) 193-297.}

\def\GVi{[10]}
\def\refGVi{ 
  R. Gopakumar and C. Vafa, {\it M-Theory and Topological Strings--I}, 
  hep-th/9809187.}

\def\GVii{[11]}
\def\refGVii{ 
  R. Gopakumar and C. Vafa, {\it M-Theory and Topological Strings--II}, 
  hep-th/9812127.}

\def\HKTY{[12]}
\def\refHKTY{ 
  S. Hosono, A. Klemm, S. Theisen, and S.-T. Yau, 
  {\it Mirror symmetry, mirror map and applications to Calabi-Yau 
  hypersurfaces},  
  Commun. Math. Phys. 167(1995), 301-350. } 

\def\HLY{[13]}
\def\refHLY{ 
  S. Hosono, B.H. Lian, and S.-T. Yau, {\it  
  GKZ-Generalized Hypergeometric Systems in Mirror Symmetry 
  of Calabi-Yau Hypersurfaces},
  Commun. Math. Phys. 182(1996), 535-577. }

\def\HSS{[14]}
\def\refHSS{ 
  S. Hosono, M.-H. Saito and J. Stienstra, 
  {\it On Mirror Symmetry Conjecture for Schoen's Calabi-Yau 3-folds}, 
  in The Proceedings of Taniguchi Symposium, ``{\it Integrable
  Systems and Algebraic Geometry}'', Kobe/Kyoto(1997), pp.194-235 
  (alg-geom/9709027). }

\def\KMV{[15]}
\def\refKMV{ 
   A. Klemm, P. Mayr and C. Vafa,
   {\it BPS States of Exceptional Non-Critical Strings}, 
   in the proceedings of the conference "Advanced Quantum Field Theory'' 
   (in memory of Claude Itzykson) hep-th/9607139. }

\def\KM{[16]}
\def\refKM{ 
 M. Kontsevich and Yu. Manin, {\it Gromov-Witten classes, quantum 
 cohomology and enumerative geometry}, Commun. Math. Phys. 164(1994), 525-562.}

\def\Ma{[17]}
\def\refMa{
    Yu. Manin, 
    {\it Generating Functions in Algebraic Geometry and Sums over Trees}, 
    alg-geom/9407005.} 

\def\MM{[18]}
\def\refMM{
 M. Marino and G. Moore, {\it  
    Counting higher genus curves in a Calabi-Yau manifold }, hep-th/9808131. }

\def\MNW{[19]}
\def\refMNW{
  J. A. Minahan, D. Nemeschansky and N. P. Warner, 
  {\it  Partition functions for the BPS states of the E8 non-critical string},
  hep-th/9707149.} 

\def\MNVW{[20]}
\def\refMNVW{ 
  J. A. Minahan, D. Nemeschansky, C. Vafa and N. P. Warner, 
  {\it  E-Strings and N=4 Topological Yang-Mills Theories},
  hep-th/9802168.} 

\def\MorI{[21]}
\def\refMorI{ 
  D. R. Morrison, {\it Picard-Fuchs equations and mirror maps 
  for hypersurfaces}, Essays on Mirror Manifolds (S.-T. Yau, ed.), 
  Internal Press, Hong Kong, (1992), 241-264. }

\def\MorII{[22]}
\def\refMorII{ D.R. Morrison, {\it Mathematical Aspects of Mirror Symmetry}, 
  in Complex Algebraic Geometry (J. Koll\'ar ed.), 
  IAS/Park City Mathematical Series, vol. 3, 1997, pp. 265-340, 
  alg-geom/9609021.}

\def\Pa{[23]}
\def\refPa{
   R. Pandharipande, {\it Hodge integrals and degenerate contributions}, 
   math.AG/9811140.}

\def\SaII{[24]}
\def\refSaII{
   M.-H. Saito, 
  {\it Prepotentials of Yukawa couplings of certain Calabi-Yau 3-folds 
   and mirror symmetry}, to appear in Proceedings of the 1998 NATO-ASI/CRM 
   Summer School on the Arithmetic and Geometry of Algebraic Cycles. . }

\def\Shi{[25]}
\def\refShi{
   T. Shioda, {\it On the Mordell-Weil lattices}, 
   Comment. Math. Univ. St. Pauli 39 (1990), 211-240.} 

\def\VW{[26]}
\def\refVW{ 
   C. Vafa and E. Witten, {\it A strong coupling test of S-duality}, 
   Nucl. Phys. B431 (1994) 3-77.}

\def\YZ{[27]}
\def\refYZ{ 
  S.-T. Yau and E. Zaslow, {\it BPS states, string duality, and nodal curves 
  on K3}, Nucl. Phys. B471 (1996) 503-512.}

\def\Zagier{[28]}
\def\refZagier{
  D. Zagier, {\it A Modular Identity Arising From Mirror Symmetry },
  to appear in The Proceedings of Taniguchi Symposium, ``{\it Integrable
  Systems and Algebraic Geometry}'', Kobe/Kyoto(1997), pp.477-480. }


\rightline{January, 1999}
\rightline{ Published form in Adv. Theor. Math. 3 (1999) 177--208}
\vskip0.5cm
\centerline{\bf Holomorphic Anomaly Equation and BPS State Counting }
\centerline{\bf of Rational Elliptic Surface}

\vskip1cm

\centerline{
S. Hosono$^1$, 
\footnote{}{\hskip-0.8cm email:
$^1$  hosono@ms.u-tokyo.ac.jp,  
$^2$  mhsaito@math.kobe-u.ac.jp, 
$^3$  atsushi@kurims.kyoto-u.ac.jp } 
M.-H. Saito$^2$, 
and 
A. Takahashi$^3$ } 

\vskip0.5cm

\bigskip\centerline{
\vbox{
\hbox{ $^1$ Graduate School of}
\hbox{ $\;\;$ Mathematical Sciences}
\hbox{ $\;\;$ University of Tokyo}
\hbox{ $\;\;$ Komaba 3-8-1, Meguro-ku  } 
\hbox{ $\;\;$ Tokyo 153-8914  }
\hbox{ $\;\;$ Japan }  }
\hskip0cm
\vbox{
\hbox{ $^2$ Department of Mathematics}
\hbox{ $\;\;$ Kobe University}
\hbox{ $\;\;$ Rokko  }
\hbox{ $\;\;$ Kobe  657-8501}     
\hbox{ $\;\;$ Japan }
\hbox{ }  } 
\hskip0cm
\vbox{
\hbox{ $^3$ Research Institute for}
\hbox{ $\;\;$ Mathematical Sciences}
\hbox{ $\;\;$ Kyoto University}
\hbox{ $\;\;$ Kyoto 606-8502 } 
\hbox{ $\;\;$ Japan }
\hbox{ }    }  }

\vskip1cm
 
We consider the generating function (prepotential) for Gromov-Witten 
invariants of rational elliptic surface. We apply the local mirror 
principle to calculate the prepotential and prove a certain recursion 
relation, holomorphic anomaly equation, for genus 0 and 1. We propose 
the holomorphic anomaly equation for all genera and apply it to 
determine higher genus Gromov-Witten invariants and also the 
BPS states on the surface. Generalizing G\"ottsche's 
formula for the Hilbert scheme of $g$ points on a surface, we 
find precise agreement of our results with the proposal recently 
made by Gopakumar and Vafa{\GVii}.

\vskip1cm

\newsec{ Introduction } 

Since the pioneering work by Candelas et al. in 1991{\CdGP}, 
the theory of the Gromov-Witten invariants has been one of the 
central topics in mathematical physics related to string theory. 
Due to many contributions on this subject we have now well-developed 
mathematical theory{\KM}{\BM} of the invariants as well as the concrete 
methods to calculate them applying the mirror symmetry of Calabi-Yau 
manifolds. However up to very recently our concrete methods 
have been restricted to the genus zero or genus one Gromov-Witten invariants. 
Although we have mathematical definition 
of the higher genus Gromov-Witten invariants, little was known about how 
to determine them explicitly for a given Calabi-Yau manifold. 
Regarding this a breakthrough 
has been made recently in {\MM} for a special class of Calabi-Yau manifolds 
which have a K3 fibration and have a dual description in the heterotic string. 
Independently Gopakumar and Vafa{\GVi}{\GVii} have derived a general form 
of the prepotential for the higher genus Gromov-Witten invariants, 
which includes several 
interesting mathematical predictions on the Gromov-Witten invariants. 

In this paper we will propose a recursion relation {\it holomorphic anomaly 
equation} as a basic equation for the higher genus Gromov-Witten invariants 
of rational elliptic surface, and will make explicit predictions for them. 
We find exquisite agreement of our results with those by Gopakumar and Vafa.

To state main results of this paper let us consider a generic 
rational elliptic surface obtained by blowing 
up nine base points of two generic cubics in $\IP^2$. Under the assumption 
for the cubics the surface $S$ has an elliptic fibration over $\IP^1$ with 
exactly twelve singular fibers of Kodaira $I_1$ type. 
We consider a situation in which the generic rational elliptic 
surface $S$ appears as a divisor in a Calabi-Yau 3-fold $X$. Since 
the normal bundle ${\cal N}_{X/S}$ is given by the canonical bundle 
$K_S$ we can extract the genus $g$ Gromov-Witten invariants 
$N_g(\beta)$ of class $\beta \in H_2(S,\ZZ)$ taking a suitable limit 
of the prepotential of the Calabi-Yau 3-fold $X$, which is called 
{\it local mirror principle}.  Since even for genus zero invariants 
the determination of $N_{g=0}(\beta)$ is technically tedious, 
in what follows, we will mainly be concerned with the following 
sum of the invariants 
$$
N_{g; d,n}:=\sum_{(\beta,H)=d,\; (\beta,F)=n} N_g(\beta) \;\,
$$
where $H$ and $F$ represent the pull back of the hyperplane class 
of $\IP^2$ and the fiber class, respectively. Associated to these 
invariants we define generating functions; 
$$
Z_{g;n}(q):=\sum_{d=0}^\infty N_{g; d,n} q^d \;\;,\;\;
F_g(q,p):=\sum_{n=0}^\infty Z_{g;n}p^n \;\;.
$$
The latter is the genus $g$ prepotential in topological string theory. 
For $g=0$ and $g=1$ we determine it via the local mirror principle 
applying to $X$, and find a recursion relation satisfied by $Z_{g;n} \;\; 
(g=0,1 ; n=1,2,\cdots)$ which we generalize for arbitrary $g$ as follows: 

\noindent
{\bf Conjecture 1.1} (Holomorphic anomaly equation for all $g$) 
{\it The generating function $Z_{g;n}(q)$ has the form 
\eqn\weakquasimod{
P_{2g+2n-2}(\phi,E_2,E_4,E_6) \(Z_{0;1}(q)\)^n  \quad,
}
with some 'quasi-modular form' for the modular subgroup $\Gamma(3)$ 
of weight $2g+2n-2$. 
(In the special cases of $g=0$ and $n=1$, it simplifies to 
$P_{2n-2}(E_2,E_4,E_6)$ and $P_{2g}(E_2(q^3),E_4(q^3),E_6(q^3))$, 
i.e., exactly the quasi-modular forms of weight $2n-2$ and $2g$, 
respectively). And it satisfies the recursion relation
\eqn\holoano{
{\pd Z_{g;n} \over \pd E_2} = {1\over 72} \sum_{g'+g''=g} \sum_{s=1}^{n-1} 
s(n-s) Z_{g';s} Z_{g'';n-s} +{n(n+1) \over 72 } Z_{g-1;n} \;\;.
} }

\vskip0.3cm
\noindent
We may 'integrate' our holomorphic anomaly equation under certain 
vanishing conditions.  In this paper we focus mainly on the special case 
of $n=1$ in which the equation simplifies to 
\eqn\recursGI{
{\pd Z_{g;1} \over \pd E_2} = {1\over 36} Z_{g-1;1}  \;\;.  }
We integrate {\recursGI} with the vanishing conditions 
and the initial data
$$
Z_{0;1}(q)= q^{3\over2} {\Theta_{E_8}(3t, t\gamma) \over \eta(q^3)^{12} }
$$
which has been found in {\KMV}{\HSS}. The $E_8$ theta function comes from the 
Mordell-Weil group of the rational elliptic surface and the eta functions 
in the denominator come from the twelve singular fibers. 
See {\HSS} for the details and notations. Then we find that the 
solutions $Z_{g;1}$ may be arranged into an all genus partition 
function of the topological string theory:

\noindent
{\bf Proposition 1.2} (Topological string partition function on $S$)  
{\it 
\eqn\mainII{
q^{3\over2} {\Theta_{E_8}(3t, t\gamma) \over \eta(q^3)^{12} }
\prod_{n\geq 1} {(1-q^{3n})^4 \over (1-t_L q^{3n})^2 (1-{1\over t_L} q^{3n})^2}
= \sum_{g\geq 0} 
Z_{g;1}(q) \lambda^{2g-2} (2{\rm sin} {\lambda\over2})^2\;\;, } 
where $\lambda$ represents the string coupling and 
$t_L={\rm e}^{ i \lambda }$. } 

\vskip0.3cm

We derive the same result following the proposal made in 
{\GVii} for the BPS state counting of the families of 
genus $g$ curves. From this viewpoint our result {\mainII} comes 
from the following generalization of G\"ottsche's formula{\G} for 
the Hilbert scheme $S^{[g]}$ of $g$ points on a surface $S$:

\noindent
{\bf Proposition 1.3} 
(G\"ottsche's formula with $SL(2,\IC)_L\times SL(2,\IC)_R$ Lefschetz action)  
{\it For the Hilbert scheme $S^{[g]}$ of $g$ points on a surface $S$ with 
a fibration structure we can decompose the Lefschetz $SL(2,\IC)$ action 
on $H^*(S^{[g]})$ into the product $SL(2,\IC)_L\times SL(2,\IC)_R$, one for 
the natural fiber space of $S^{[g]}$ and the other for the base space. 
If we write the Poincar\'e polynomial by
$$
P_{t_L,t_R}(S^{[g]})= (t_L t_R)^g {\rm Tr}_{H^*(S^{[g]})}
(t_L^{2 j_{3,L}} t_R^{2 j_{3,R}}) \;\;,
$$
then the generating function 
$G(t_L,t_R,q)=\sum_{g\geq 0} P_{t_L,t_R}(S^{[g]}) q^g$, for the surface 
with $b_1(S)=0$, is given by
\eqn\GLR{
\eqalign{
G(t_L,t_R,&q)=\prod_{n\geq1} \bigg\{ {1 \over 
\(1-(t_Lt_R)^{n-1}q^n\)
\(1-(t_Lt_R)^{n+1}q^n\)} \cr
& \times { 1\over 
\(1-t_L^2(t_Lt_R)^{n-1}q^n\)\(1-t_R^2(t_Lt_R)^{n-1}q^n\)
\(1-(t_Lt_R)^{n}q^n\)^{b_2(S)-2} } \bigg\} \;\;. \cr}
}    }

\vskip0.3cm

We explain our result {\mainII} in terms of the above generalization 
of G\"ottsche's formula by
\eqn\thetaG{
\Theta_{E_8}(3t,t\gamma) G(-t_L,-1,{q^3 \over t_L})
= \sum_{g\geq 0} 
Z_{g;1}(q) \lambda^{2g-2} (2{\rm sin} {\lambda\over2})^2\;\;. } 
This implies that the genus $g$ curves ${\cal C}_g$ in $S$ satisfying 
$({\cal C}_g,F)=1$ split into irreducible parts, one coming from the 
Mordell-Weil group and the others from elliptic curves (with possible nodal 
singularities) in the fiber direction. 

The readers who are not interested in the derivation and the proofs of 
the holomorphic anomaly equation may omit the following two sections and 
may start from the section 4 for our main results.

\vskip0.3cm
The organization of this paper is as follows: In section 2, we will 
introduce a Calabi-Yau hypersurface and its mirror, 
and introduce the hypergeometric series representing the 
prepotential $F_0(t)$ for the Calabi-Yau hypersurface. In section 3, 
we will take a limit to reduce the prepotential to the one relevant 
to the rational elliptic surface (, {\it local mirror principle}). 
We will analyze the reduced prepotential and 
the mirror maps in detail, and will prove the recursion relation, 
holomorphic anomaly equation at $g=0$. Using the formula in {\BCOVi} 
for $F_1$, we will also prove the recursion relation at $g=1$. 
In section 4, we will {\it propose} our recursion relation for all genera, 
and solve the recursion relation with some vanishing conditions. 
There we also discuss about the Gromov-Witten invariants 
coming from $Z_{g;1}(q)$. 
In the final section, we discuss some relations to the recent 
developments on the counting problem of the BPS states in topological 
string theory{\GVii}{\KMV}. There we will find a generalization of 
G\"ottsche's formula for the Poincar\'e polynomials of $S^{[g]}$.

\vskip0.5cm
\noindent
{\bf Acknowledgments:}
Two of the authors (S.H. and M.-H.S.) would like to thank N.Yui for 
her kind hospitality during their stay (August, 1998) at Queen's 
University where very early stage of this work has been done. 
The research of S.H. and M.-H.S. is supported in part by Grant-in Aid 
for Science Research (A-09740015 for S.H. and B-09440015 for M.-H.S.), 
the Ministry of Education, Science and Culture, Japan. The research 
of A.T. is supported by Research Fellowships of Japan Society for 
the Promotion for Young Scientists.  


After our submission of this paper to the e-print archive, hep-th, we are 
informed by A. Klemm that he is testing the higher genus prepotentials 
$F_g$ for several surfaces other than our rational elliptic surface 
(work to appear). We would like to thank him for sending us his 
preliminary draft prior to publication. 

\vskip1cm \noindent
{\bf Note added in proof:} In references [15] and [19], a different 
base $F+e_9$, in stead of our $H$, is used to define $Z_{g=0;n}$. In this 
case we consider our generating function $Z_{g;n}$ for the invariants  
$N_{g;d,n}=\sum N_g(\beta)$ summed over $\beta$ with $(\beta,F+e_9)=d, 
(\beta,F)=n$. Then our Conjecture 1.1 should be read as follows; 

{\it The generating function $Z_{g;n}(q)$ has the form 
$$\displaystyle{
P_{2g+6n-2}(E_2,E_4,E_6) { q^{n\over 2} \over \eta(q)^{12n} }  
}
$$
with a quasi-modular form $P_{2g+6n-2}(E_2,E_4,E_6)$ of weight $2g+6n-2$, 
and satisfies the same holomorphic anomaly equation as 1.2 replaced by  
the prefactors ${1 \over 72}$ and ${1 \over 24}$. }

It is worth while remarking here that in this case 
the integration constants may be determined {\it consistently} for all 
$g$ and $n$ by simply requiring the vanishing conditions for the first 
few terms in the $q$-expansion of $\tilde Z_{g;n}(q)$, 
which is defined by 4.5 (, 3.18 and 3.27) with 
$D(g,h,k)=C_h(g-h,1) k^{2g-3} \; (0\leq h \leq g)$[11].

\newsec{Mirror symmetry of a Calabi-Yau hypersurface in $\IP^2 \times \IF_1$ }
\global\global\propno=1

In this section we will consider a Calabi-Yau hypersurface which 
contains a generic rational elliptic surface as a divisor. We collect 
necessary formulas for the (genus zero) prepotential. 

Let us start with the Hirzebruch surface $\IF_1$ which is defined by 
the quotient $(\IC^4\setminus {\cal Z}) /\sim$ with 
$$
{\cal Z}=\{ (u_1,u_2,u_3,u_4)\in \IC^4 \;\vert\; 
             u_1=u_2=0 \; {\rm or} \; u_3=u_4=0 \} 
$$
and $\IC^*$-actions 
$$
(u_1,u_2,u_3,u_4) 
\sim (\lambda_1 u_1,\lambda_1 u_2,1/\lambda_1 u_3,u_4) 
\sim (u_1, u_2, \lambda_2 u_3, \lambda_2  u_4)  \;\;,\;
(\lambda_1, \lambda_2 \in \IC^*)\;\;.
$$
We may consider a generic hypersurface in the product of the surface 
$\IF_1$ with $\IP^2$ given by the data 
$$ X= \cicy{}[\IP^2 \cr \IF_1 \cr | 3 \cr (1,2) \cr ]{3,75} $$
where $(1,2)$ refers to the homogeneous degrees with respect to 
the first scaling by $\lambda_1$ and the second one by $\lambda_2$. 
The defining equation may be written explicitly as
\eqn\defW{
g_{3,3}(z_1,z_2,z_3,u_1,u_2)u_3^2 + f_{3,1}(z_1,z_2,z_3,u_1,u_2)u_4^2=0\;\;,
}
where $z_1,z_2,z_3$ represents the homogeneous coordinate of $\IP^2$ and 
$g_{3,3} \;(, f_{3,1}, )$ 
refers to a generic homogeneous polynomial with bi-degree 
$(3,3) \;(,$ and $(3,1)$, respectively,) for the coordinates 
$z_1,z_2,z_3$ and $u_1,u_2$. 
This is an elliptic Calabi-Yau hypersurface over $\IF_1$ with 
the Hodge numbers $h^{1,1}=3$ and $h^{2,1}=75$. Two of the three 
elements in $H^{1,1}(X)$ come from the base $\IF_1$ and the other 
comes from the fiber elliptic curve. We may find in $X$ 
a rational elliptic surface $S$ with its defining equation of bi-degree 
$(3,1)$ in $\IP^2\times\IP^1$.  It appears as a divisor $u_3=0$, 
which is the cubics in $\IP^2$ over the $(-1)$ curve in $\IF_1$.

The positive classes in $H^2(X,\ZZ)$ are generated by the three 
integral elements in $H^{1,1}(X)$ corresponding to the divisors 
$$
H=(z_1=0)\cap X \;\;,\;\; F=(u_1=0)\cap X \;\;,\;\; D=(u_4=0)\cap X \;\;. 
$$
We sometimes denote the corresponding forms by $J_1, J_2 $ and $J_3$, 
respectively. 
It is straightforward to determine the non-zero intersection numbers 
$K_{abc}^{top}=\int_X J_a\wedge J_b \wedge J_c$ and 
$c_2J_a = \int_X c_2(X) \wedge J_a$ 
with the second Chern class $c_2(X)$ to be
$$
\eqalign{ &
K^{top}_{112}=2 \;\;,\;\; 
K^{top}_{113}=3 \;\;,\;\; 
K^{top}_{123}=3 \;\;,\;\; 
K^{top}_{133}=3 \;\;, \cr 
& c_2J_1=36 \;\;,\;\; c_2J_2=24 \;\;,\;\; c_2J_3=36 \;\; .\cr}
$$

The ambient space $\IP^2\times \IF_1$ is so-called the toric Fano manifold, 
and thus we can easily construct the mirror Calabi-Yau hypersurface $X^\vee$  
based on Batyrev's toric method[Bat].  Furthermore the prepotential 
of the mirror Calabi-Yau hypersurface $X^\vee$ is determined by the general 
formula obtained in {\HKTY}{\HLY}.  

Here we collect necessary formulas to determine the prepotential. 
We start with a hypergeometric series representing a period integral 
for a deformation family of $X^\vee$ parameterized locally by $x,y,z$; 
$$
w_0(\vec x)=\sum_{n,m,k\geq0} c(n,m,k) x^n y^m z^k \;\;, 
$$
with 
$$
c(n,m,k)={\Gamma(1+3n+m+2k) \over 
\Gamma(1+n)^3 \Gamma(1+m)^2 \Gamma(1+k-m) \Gamma(1+k)} \;\;.
$$
The local parameters $(x,y,z)$ has been chosen so that its origin represents 
the celebrated boundary point where the monodromy is maximally 
degenerated{\MorI}{\MorII}. The series $w_0(\vec x)$ represents the period 
integral for the invariant cycle about this degeneration point and 
satisfies Picard-Fuchs differential equation (see Appendix). 
As a complete set of the solutions of the Picard-Fuchs equation, we have
$\{ w_0(\vec x) \;,\; w^{(1)}_a(\vec x) \;,\; 
    w^{(2)}_b(\vec x) \;,\; w^{(3)}(\vec x) \}\; 
(a,b=1,2,3)$ where 
$$
\eqalign{
&
w^{(1)}_a(\vec x) = 
{1\over 2\pi i} {\pd \over \pd \rho_a} 
w_0(\vec x, \vec \rho)\vert_{\vec\rho=0} \;,\; 
w^{(2)}_b(\vec x) = {1\over (2\pi i)^2} {1\over 2!} 
\sum_{c,d}K^{top}_{bcd} 
{\pd \over \pd\rho_c}{\pd \over \pd \rho_d} 
w_0(\vec x, \vec \rho)\vert_{\vec\rho=0} \;,\;
\cr
&
w^{(3)}(\vec x) = -{1\over (2\pi i)^3} {1\over 3!} \sum_{a,b,c}K^{top}_{abc} 
{\pd \over \pd\rho_a}{\pd \over  \pd \rho_b}{\pd \over \pd \rho_c} 
w_0(\vec x, \vec \rho)\vert_{\vec\rho=0} \;\;
\cr}
$$
with $w_0(\vec x,\vec\rho)=\sum_{n,m,k\geq0} c(n+\rho_1,m+\rho_2,k+\rho_3) 
x^{n+\rho_1}y^{m+\rho_2}z^{k+\rho_3}$. 
In terms of the solutions of the Picard-Fuchs equation, the mirror map is 
defined by the relation
\eqn\mirrormap{
t_a = {w^{(1)}_a(\vec x) \over w_0(\vec x) } \;\; (a=1,2,3),
}
which connects the deformation parameters $(x,y,z)$ to those $(t_1,t_2,t_3)$ 
parameterizing the complexified K\"ahler moduli of $X$ at the large radius 
$({\rm Im}(t_a) \rightarrow \infty)$. Now the prepotential of 
the mirror $X^\vee$ is defined to be
$$
F(\vec x)={1\over2}{1\over w_0(\vec x)^2} 
\( w_0 \big(w^{(3)}-\sum_b {c_2J_b \over 12} w^{(1)}_b \big) 
   + \sum_a w^{(1)}_a w^{(2)}_a \) \;\;.
$$
Then the mirror symmetry conjecture 
asserts that the prepotential $F(\vec x)$ of the mirror $X^\vee$ combined 
with the mirror map provides, up to terms of classical topological 
invariants, the generating function $F_0(t)$ of the Gromov-Witten 
invariants of $X$;
$$
F(t)={1\over 3!}\sum_{a,b,c} K^{top}_{abc}t_at_bt_c
-\sum {(c_2J_b) t_b \over 24}
+{\zeta(3) \chi(X) \over 2 (2\pi i)^3} 
+{1 \over (2\pi i)^3} \sum_{0\not=\beta \in H_2(X,\ZZ)} N(\beta) 
{\rm e}^{2\pi i(\beta,\sum J_c t_c)},
$$
where we substitute the inverse relation of {\mirrormap} into $F(\vec x)$.

\vskip2cm

\newsec{Holomorphic anomaly equations for $g=0,1$}
\global\global\propno=1

\subsec{Local mirror principle and the reduction to $S$ }
Let us consider the following limit
$$
F(q,p):= F(t)-(topological \; terms) \vert_{{\rm Im t_3 \rightarrow \infty}}
$$
with $q={\rm e}^{2\pi i t_1}$ and $p={\rm e}^{2\pi i t_2}$. 
Since the class $J_3$ measures the volume of the fiber $\IP^1$ of $\IF_1$ 
parameterized by $u_3,u_4$ and 
the volume of the curve contained in the divisor $S=(u_3=0)\cap X$ 
are measured to be zero by this class, 
the limit ${\rm Im}t_3 \rightarrow \infty$ throw away all 
the Gromov-Witten invariants except those of the curves contained in the 
rational elliptic surface $S$.  Thus we may expect that the reduced 
prepotential $F(q,p)$ coincides with the generating function defined 
in Section 1. This is so-called the local mirror principle, and somehow 
generalize the arguments done for the isolated 
$(-1,-1)$-curves in Calabi-Yau manifolds{\Ma}{\FP}. 

In our case of the elliptic Calabi-Yau hypersurface $X$ 
in $\IP^2\times\IF_1$, the limit 
${\rm Im}t_3 \rightarrow \infty$ which translates to the limit 
$z\rightarrow 0$ in the mirror $X^\vee$ greatly simplifies 
the period integral $w_0(\vec x)$; 
$$
\eqalign{
w_0(x,y,z)\vert_{z=0} 
&= \sum_{n\geq0}{\Gamma(1+3n) \over \Gamma(1+n)^3} x^n \;\;, \cr
&=: \phi(x)  \;\;.}
$$
We note that the series $\phi(x)$ is nothing but a solution of 
the Picard-Fuchs equation of the fiber elliptic curve of $S$:
\eqn\PFell{
\{\theta_x^2-3x(3\theta_x+2)(3\theta_x+1)\}\phi(x) = 0  \;\;.
}
Another solution of {\PFell} about $x=0$ may be given by 
$$
\tilde \phi(x)={\rm log}(x) \phi(x) + 
\sum_{n\geq0}{\Gamma(1+3n) \over \Gamma(1+n)^3}
\( \psi(1+3n) - 3 \psi(1+n) \) x^n \;\;,
$$
where $\psi(z)={d \over d z} {\rm log}\Gamma(z)$. 

Now looking at the relations between the 'periods' of $X$ via the 
prepotential and the period integrals of $X^\vee$:
$$
\(1,t_a, {\pd F \over \pd t_b }, 2F-\sum_c t_c {\pd F \over \pd t_c} \) 
= {1\over w^{(0)}}\( w^{(0)}, w^{(1)}_a, 
w^{(2)}_b, w^{(3)}-{1\over12}{\sum_c  (c_2J_c) w^{(1)}_c} \) \;,
$$
it is straightforword, although involved technically, to derive the 
following consice form for the derivative 
${\pd \over \pd t_p}F(p,q) \;\;(t_p:=t_2) $ 
under the limit ${\rm Im}t_3 \rightarrow \infty$ (cf. Sect.7 of {\HSS}).

\vskip1cm

\noindent
{\bf Proposition }\prop\PropFp 
{\it 
$$
\pdp F(q,p) = \sum_{n\geq1} {f_n(x) \over \phi(x)^2} y^n \;\;,
$$
where we define
$$
f_n(x)=-{1\over3}
\{\phi(x){\cal L}_n\tilde \phi(x) - \tilde \phi(x) {\cal L}_n \phi(x) \}\;,
$$
with a linear operator 
$$
{\cal L}_n={(-1)^n  \over n \times n!} \prod_{k=1}^n (3\theta_x+k) \;\;.
$$
}

\noindent
{\bf Remark }\prop\RemEll
By constructing Barnes integral representation of the series $\phi(x)$, 
it is an easy exercise to make an integral symplectic basis of 
the Picard-Fuchs equation {\PFell}. It turns out that our bases 
$\phi(x)$ and $\tilde\phi(x)$ in fact constitute an integral symplectic 
basis about $x=0$. Therefore we may write the 
holomorphic one form of the elliptic curve by $\Omega_E=\phi_0 A + \phi_1 B$, 
where $A$ and $B$ are symplectic bases of the elliptic curve. 
Then the function $f_1(x)$ may be written by 
$-3 \int_E \Omega_E \wedge \theta_x \Omega_E$. 
This is so-called the classical Yukawa coupling {\CdGP} 
of the elliptic curve, and may be determined from the Picard-Fuchs 
equation {\PFell} to be $f_1(x)={9 \over 1-27 x}$. For the other 
functions $f_m(x) (m\geq 2)$ we will find a powerful recursion relation.

\vskip0.5cm

The relation {\mirrormap} is also simplified in the limit $z\rightarrow 0$ 
due to the following relations
$$
w_0(x,y,0)=\phi(x)\;\;,\;\;w^{(1)}_1(x,y,0)=\tilde \phi(x) \;\;,\;\;
w^{(1)}_2(x,y,0)=\xi(x)+\sum_{m\geq1}{\cal L}_m \phi(x) y^m \;\;,
$$
where $\xi(x)=\sum_{n\geq0}{(3n)!\over (n!)^3}\( \psi(3n+1)-\psi(1) \) x^n$.
As the inverse relations of {\mirrormap}, we have $x=x(q,p)$ and $y=y(q,p)$ 
with $q={\rm e}^{2\pi i {w^{(1)}_1(x,y,0) \over w_0(x,y,0)} }$ and  
$p={\rm e}^{2\pi i {w^{(1)}_2(x,y,0) \over w_0(x,y,0)} }$. 

\vskip1cm
\noindent
{\bf Proposition} \prop\PropMirrormap
{\it 
Under the limit $z\rightarrow0$, we find that: 
\item{(1)} The inverse series $x(q,p)$ does not depend on $p$ and is given 
by the level three modular function;
\eqn\xmirror{
x(q,p)={1\over t_{3B}(q)} \;\;,
}
where $t_{3B}(q)={\eta^{12}(q) \over \eta^{12}(q^3)}$ is the Thompson series 
in the notation of {\CN}. 
\item{(2)}
The inverse series $y(q,p)$ is determined iteratively as a power series 
of $p$ through the relation
\eqn\ymirror{
y(q,p)=p\psi(x){\rm e}^{-\sum_{m\geq1}c_m(x)y^m}\;\;,
}
where $\psi(x)={\rm e}^{-{\xi(x) \over \phi(x)}}$ and 
$c_m(x)={{\cal L}_m \phi(x) \over \phi(x)}$. 
\par}

\vskip1cm
\noindent
{\bf Remark} \prop\RemGS
The function $\psi(x)$ with $x={1\over t_{3B}(q)}$ has first 
appeared in {\HSS}, and has been determined in terms of the modular 
functions of level three{\Zagier};
$$
\psi(x(q))=q^{{1\over6}}\(t_{3A}(q)\)^{-{1\over2}}\(t_{3B}(q)\)^{2\over3}\;\;,
$$
where $t_{3A}(q)={1 \over x(1-27x) }$.
Also the following relations are standard results coming from the 
Gauss-Schwarz theory for the Picard-Fuchs equation {\PFell};
\eqn\phiq{
\phi(x(q))=\theta_3(q)\theta_3(q^3)+\theta_2(q)\theta_2(q^3)  \;\;,\
}
\eqn\modularphi{
\phi(x)^{12}x(1-27x)^3=\eta(q)^{24} \;\;,\;\;
{1\over2\pi i} {d x \over d t} = \phi(x)^2 x (1-27x) \;, 
}
where $\theta_2(q)=\sum_{m\in \ZZ} q^{(m+{1\over2})^2}$ and 
$\theta_3(q)=\sum_{m\in \ZZ} {q}^{m^2}$.

\vskip1cm
The following lemma may be derived directly from the 
definition $c_m(x)$ and the relation 
$$
{\theta_x \phi(x) \over \phi(x)} = -{1\over3}\(
1-{f_1\over12}-{f_1\over36}{E_2(q) \over \phi(x)^2} \) \;,
$$
which follows from {\modularphi};

\vskip0.5cm
\noindent
{\bf Lemma} \prop\LemmaRec
{\it 
Under the relation $x={1\over t_{3B}(q)}$, the function 
$c_m(x)={{\cal L}_m \phi(x) \over \phi(x)}$ may be written by 
\eqn\cmx{
c_m(x)=B_m(f_1){E_2(q) \over \phi(x)^2} + D_m(f_1) \;\;,
}
where $B_m$ and $D_m$ are some polynomials of $f_1$ determined 
by the following recursion relation;
\eqn\recursBD{
\eqalign{ 
B_{m+1}& =-{m\over (m+1)^2} \left\{ 
(3\theta_x+m+2-{f_1 \over 12})B_m + {f_1 \over 36} D_m \right\} \;\;, \cr 
D_{m+1}&=-{m\over(m+1)^2} \left\{ 
-{1\over 4}(f_1-8)B_{m}+(3\theta_x+m+{f_1\over12})D_m \right\}  \;, \cr }}
with initial values $B_1=-{f_1 \over 36}$ and $D_1=-{f_1 \over 12}$. 
}

\vskip1cm
We present here the first few terms coming from the recursion 
relation {\recursBD};
\eqn\Bms{
B_2={1\over432}f_1^2 \;\;,\;\; 
B_3={7\over5832}\(f_1^2 -{2\over7} f_1^3 \)  \;\;,\;\;
B_4={1\over3888}\(f_1^2-{5\over3}f_1^3+{1\over4}f_1^4\) \;, \cdots  }
\eqn\Dms{
\eqalign{
 D_2=-{1 \over 36}& \(f_1-{f_1^2 \over 4}\) \;\;,\;\;
  D_3=-{1\over 162}\(f_1-{5f_1^2 \over 4}+{f_1^3 \over 6}\) \;\;,  \cr
& 
  D_4={17\over3888}\(f_1^2-{8f_1^3\over17}+{3f_1^4\over68}\)\;, \cdots  }}
We can verify directly using the Picard-Fuchs equation {\PFell} that 
the formal solution of the recursion relation may be written  
in terms of the functions $f_m(x)$ in Proposition {\PropFp} as 
\eqn\formalsolBD{
B_m=-{f_m \over 36} \;\;,\;\; 
D_m={1\over f_1} \left\{ {(m+1)^2 \over m} f_{m+1} + 
 (3\theta_x+m+2-{f_1 \over 12}) f_m \right\} \;\;. }
As a result we see that the functions $f_m(x)$'s may be determined 
in terms of the recursion relation {\recursBD}. 

Since both the $B_m$ and $D_m$ are polynomials of $f_1(x)={9\over 1-27x} =
9 {t_{3A}(q) \over t_{3B}(q)}$, they have nice behavior under the level 
three modular subgroup $\Gamma(3)$. Therefore the {\it modular anomaly} 
comes from the $E_2$-term in $c_m$. We may express this anomalous behavior 
via the partial derivative of $c_m$;
\eqn\Dcm{
{\pd c_m(x) \over \pd E_2} = -{1 \over 36}{f_m(x) \over \phi(x)^2} \;\;, }
which plays a central role in the following derivations of the holomorphic 
anomaly equations.

\subsec{ Holomorphic anomaly equation at $g=0$ }

Now we are ready to prove the recursion relation for $Z_{0;n}(q)$'s, 
which come from the mirror symmetry conjecture through the expansion 
\eqn\BmodelZn{
\pdp F(q,p) = \sum_{n\geq1} {f_n(x) \over \phi(x)^2} y^n 
= \sum_{n\geq 1} n Z_{0;n}(q) p^n \;\;.
}

\vskip1cm

\noindent
{\bf Theorem} \prop\ThmRecZero (Holomorphic anomaly equation at $g=0$ 
(c.f.{\MNW})) 
{\it The function $Z_{0;n}$ satisfies the recursion relation 
\eqn\recZn{
{\pd Z_{0;n} \over \pd E_2 } = 
{1 \over 72}\sum_{s=1}^{n-1} (n-s)s Z_{0;n-s} Z_{0;s} \;\;\;\;(n\geq1).} }
\par\noindent
(Proof) 
{}From Proposition {\PropFp} and Proposition {\PropMirrormap}, we have
$$
\pdp F(q,p)=
\sum_{n=1}^\infty {f_n(x) \over \phi(x)^2} y^n \;\;\;,\;\;\; 
y = p \psi(x) {\rm e}^{-\sum_{m\geq1}c_m y^m} \;\;, 
$$
where the quasi-modular property (anomalous modular property) appears 
in $c_m$ through {\Dcm}. 
Now we first note 
$$
\pdp y = {y \over 1+\sum_{m=1}^\infty m c_m y^m } \;\;\;.\;\;\;
$$
Using this and the relation {\Dcm}, we have 
\eqn\yE{
{\pd y \over \pd E_2} = 
{-y \sum_{m=1}^\infty {\pd c_m \over \pd E_2} y^m 
  \over 
 1+\sum_{m=1}^\infty m c_m y^m } 
={1\over 36} (\pdp y)(\pdp F)  
\;\;. }
Now we have
$$
\eqalign{
{\pd \over \pd E_2} (\pdp F) 
&= {1\over \phi^2}\sum_{m\geq1} f_m(x) m y^{m-1} {\pd y \over \pd E_2} \cr
&= {1\over \phi^2}\sum_{m\geq1} f_m(x) m y^{m-1} {1\over 36}(\pdp y)(\pdp F) \cr
&= {1\over 36}((\pdp)^2F)(\pdp F) \;\;, \cr}
$$
which says, up to constant terms for $p$, that
$$
{\pd F(q,p) \over \pd E_2} = {1\over 72} \( \pdp F(q,p) \)^2 \;\;.
$$
This proves the recursion relation. \hfill []

\vskip1cm

Now we determine the explicit forms $Z_{0;n}$ for lower $n$ from 
the formula {\BmodelZn}. After coming to a conjecture about the 
form of $Z_{0;n}$, we remark that the holomorphic anomaly equation 
{\recZn} with some obvious inputs for the Gromov-Witten invariants 
suffices to determine the form $Z_{0;n}$ for all $n$. 

Since our formula {\BmodelZn} is written in terms of the known 
functions $f_1(x), \psi(x)$ with $x={1\over t_{3B}(q)}$ and $E_2(q)$ 
for each order of $p$, and also the order of $p$ coincides with the order 
of $\psi(x)$, it is easy to deduce that $Z_{0;n}$ has the 
following form in general, 
\eqn\Gnf{
Z_{0;n}=G_{0;n}(f_1, {E_2 \over \phi^2}) 
        \phi^{2n} \( {f_1 \psi \over \phi^2} \)^n  \;\;,
}
where we have factored the form of $Z_{0;1}={f_1 \psi \over \phi^2}$.
Looking into the detail of the expansion of {\BmodelZn}, we see 
in general that $\phi^2 \times G_{0;n}(f_1,{E_2\over \phi^2})$ is 
a polynomial of ${E_2 \over \phi^2}$ with 
coefficients being polynomials of $1/f_1$ over $\IQ$. Here we 
present the first few of them,
$$ 
\eqalign{
&
G_{0;2}={1\over 72} E_2 \phi^{-4} \;\;,\;\;
G_{0;3}={5\over 7776}\( (1-{8 \over f_1})\phi^4 +{3\over5}E_2^2  \)
\phi^{-6}   \cr
&
G_{0;4}=-{1\over 31104} \( (1-{12\over f_1}+{24\over f_1^2})\phi^6 
-{5\over3}(1-{8\over f_1})\phi^4 E_2-{4\over9} E_2^3 \) \phi^{-8} \cr
&
G_{0;5}=
{269 \over 62208000} \bigg(
(1-{16 \over f_1}+{64 \over f_1^2})\phi^8 + 
{6250 \over 7263}(1-{8\over f_1}) E_2^2 \phi^4 \cr
& \hskip3cm
-{2000 \over 2421}(1-{12\over f_1}+{24\over f_1^2})E_2 \phi^2 +
{3125 \over 21789} E_2^4 \bigg) \phi^{-10} \;\;.\cr}
$$
Now we note the following relations for the polynomials 
of $1/f_1$;
\eqn\fEs{
\eqalign{
E_4= & 9\(8-{8\over f_1}\)\phi^4 \;\;,\;\; 
E_6=-27\(1-{12\over f_1}+{24 \over f_1^2} \)\phi^6  \;\;, \cr
&
27 \phi^8-18E_4 \phi^4-E_4^2-8E_6 \phi^2 =0 \;\;.\cr} }
Using these relations we find that
$$
\eqalign{
G_{0;3}=&{5\over 69984}\(E_4+{27\over5}E_2^2\) \phi^{-6} \;\;,\;\;
G_{0;4}={1 \over  839808}\(E_6+5E_4 E_2+12 E_2^3 \) \phi^{-8} \;\;, \cr
& G_{0;5}={1\over 40310784} \( {269 \over 125}E_4^2+{16\over3}E_6E_2 
    +{50\over3}E_4E_2^2+25E_2^4 \)\phi^{-10} \;\;. \cr}
$$
Here we observe explicitly that $\phi$ 
disappears nontrivially in the final expression of $Z_{0;n}$ for 
lower $n$'s. We do not have general proof about this but may state 
it as follows;

\vskip0.5cm
\noindent
{\bf Conjecture }\prop\ConjZero
{\it 
The function $Z_{0;n}(q)$ in {\Gnf} takes the form
\eqn\ZnP{
Z_{0;n}(q) = P_{2n-2}(E_2,E_4,E_6) \(Z_{0;1}(q)\)^n \;\;,}
where $P_{2n-2}$ is a quasi-modular form of weight $2n-2$. 
}

\vskip1cm

\noindent
{\bf Remark} \prop\RemMulti
The function $Z_{0;n}$ contains the multiple cover contributions. 
We may subtract these contributions considering
\eqn\Znsub{
\tilde Z_{0;n}(q):=Z_{0;n}(q) - \sum_{k | n, k\not=1} 
{1\over k^3}\tilde Z_{0;n/k}(q^k) \;\;. }
The $q$ series coefficients of $\tilde Z_{0;n}$ ``count'' the 
numbers of rational curves ${\cal C}$ in our rational 
elliptic surface $S$ satisfying 
$({\cal C},H)=d$ and $({\cal C},F)=n$. 
The homology classes of curves in $S$, in general, have the form 
$[{\cal C}]=d H -a_1 e_1 -a_2 e_2 - \cdots -a_9 e_9$ 
with $a_1,a_2,\cdots,a_9 \geq 0$. 
Therefore we have $3d-a_1-a_2-\cdots-a_9 =n $
for $({\cal C},F)=n$, which implies $3d \geq n$. 
In other words, we should have 
\eqn\inidata{
\tilde N_{0;d,n}=0 \;\; {\rm for}  \;\; d < {n \over 3}  \;\;}
for the coefficients of $\tilde Z_{0;n}$. 
{}From a simple counting of the dimensionality of the quasi-modular forms 
of weight $2n-2$, we see that the vanishing condition {\inidata} 
together with the above Conjecture {\ConjZero} provides sufficient data 
to determine the integration constants for the recursion {\recZn}, 
and determine completely $Z_{0;n}$ for all $n$ (, see Section 4). 

\vskip1.5cm

\subsec{ Holomorphic anomaly equation at $g=1$ }

According to {\BCOVi}, we have an explicit expression for the 
genus one prepotential $F_1^{BCOV}(t)$ in terms of the 
discriminant of the hypersurface $X$;
\eqn\FBCOV{
F_1^{BCOV}(t)=
{\rm log} \bigg\{ \({1\over w_0(\vec x)}\)^{3+h^{1,1}(x)-{\chi(X)\over12}}
dis(x,y,z)^{-{1\over6}} 
x^{-4}y^{-3}z^{-4} {\rm det}\({\pd x_a \over \pd t_b} \) \bigg\} \;,}
where the discriminant may be determined from the characteristic variety 
of the Picard-Fuchs equation presented in Appendix. Several exponents in 
{\FBCOV} have been fixed by the requirements of the asymptotics of 
$F_1^{BCOV}$ when ${\rm Im}t_a \rightarrow \infty$. 
The explicit form of the discriminant 
is a complicated polynomial of $x,y$ and $z$, however, in the 
limit $z\rightarrow 0$, it simplifies to 
\eqn\discrim{
dis(x,y,z)\vert_{z=0}=(1-27x)^3\{1-27x+(1+y)^3-1\} \;\;.}
Also under this limit, it is easy to show from the definition 
that the mirror map $z(q,p,r) \;\;(r={\rm e}^{2\pi i t_3})$ simplifies to 
\eqn\mirrormapz{
\eqalign{
z(q,p,r)&= r {\rm e}^{-{2\xi(x) \over \phi(x)}+\sum_{m\geq1}c_m(x) y^m} \cr
        &= r \psi(x)^3 {p \over y(q,p)} \;\;.\cr }}
Now using the relations {\modularphi} it is straightforward to derive;
\vskip0.5cm

\noindent
{\bf Proposition} \prop\PropBCOV 
{\it 
When  $r\rightarrow 0$, up to the topological term 
$-{1\over12}\sum_c (c_2J_c)t_c=-3 \log q -2 \log p - 3 \log r$, we have; 
\eqn\FBCOV{
\eqalign{
F_1^{BCOV}(q,p) =&\log \big\{ 
\phi^4(1-27x)^{3\over2}q^{5\over3} 
\eta(q)^{-40} \big\}\cr
&+
\log \big\{ 
\{(1-27x)+(1+y)^3-1\}^{-{1\over6}}
{{\rm e}^{-\sum_{m\geq1}c_m(x)y^m } 
\over 1+\sum_{m\geq1}m c_m(x) y^m } \big\} \;. \cr}
} }

\vskip0.5cm
\noindent
{\bf Remark} \prop\RemNorm
\item{1)} For the $p$-independent term of $F_1^{BCOV}(q,p)$ the local 
mirror symmetry does not apply. This is because these curves are parallel to 
the fiber elliptic curves of $X$ and therefore can move 
outside of the rational elliptic surface. In fact we see that 
$$
\tilde N_{g=1;d,0} =\cases{ 4 & $d=3$ \cr 0 & $d\not=3$ \cr} 
$$
after subtracting the genus zero contributions $\tilde N_{g=0;d,0}=168 \;
(d\equiv 1,2 \; {\rm mod}\; 3) ,\; 144 \; (d\equiv 0 \; {\rm mod} \; 3)$. 
The number $4$ should be regarded as the Euler number of the base 
$\IF_1$ for the elliptic fibration. 
\item{2)} 
There is a difference in the normalization of the prepotentials 
between $F_1^{BCOV}$ and our $F_{g=1}$ in the introduction. 
These are related by the factor 2 coming from the orientation 
of curves as 
\eqn\relationF{
F_{1}(q,p)={1\over2}F_1^{BCOV}(q,p) \;\;.
}

\vskip1cm
\noindent
Now we define the generating function $Z_{1;n}$ through
$$
F_1(q,p)=\sum_{p\geq 1}Z_{1;n}(q) p^n  \;\; ,
$$
and prove the holomorphic anomaly equation at $g=1$.

\vskip0.5cm
\noindent
{\bf Theorem} \prop\ThmRecI 
(Holomorphic anomaly equation at $g=1$) 
{\it The function 
$Z_{1;n}(q)$ satisfies the following recursion relation; 
\eqn\recZngI{
{\pd Z_{1;n} \over \pd E_2 } = 
{1 \over 36}\sum_{s=1}^{n-1} (n-s)s Z_{1;n-s} Z_{0;s} 
+ {n(n+1)\over 72} Z_{0;n}\;\;(n\geq1). } 
}
\par\noindent
(Proof) Since we have already shown the relations 
$$
{\pd y \over \pd E_2}={1\over36} ({ \pdp y})( \pdp F_{0} ) \;\;, \;\;
{\pd c_m \over \pd E_2}=-{1\over36}{f_m \over \phi^2} \;,
$$
and 
$$
\pdp y = {y \over 1+\sum_{m=1}^\infty m c_m y^m } \;\;\;,\;\;\;
$$
it is straightforward to derive 
$$
\eqalign{
{\pd F_1^{BCOV} \over \pd E_2} &= 
{\pd \over \pd y} 
\log \big\{ 
\{(1-27x)+(1+y)^3-1\}^{-{1\over6}}
{{\rm e}^{-\sum_{m\geq1}c_m(x)y^m } 
\over 1+\sum_{m\geq1}m c_m(x) y^m } \big\}
{\pd y \over \pd E_2}   \cr
& \hskip2.5cm
-\sum_{m\geq1}{\pd c_m \over \pd E_2} y^m  
-{\sum_{m\geq1} m {\pd c_m \over \pd E_2}y^m \over 
  (1+\sum_{m\geq1}mc_m y^m) } \cr
&
=\pdp \log \big\{ 
\{(1-27x)+(1+y)^3-1\}^{-{1\over6}}
{{\rm e}^{-\sum_{m\geq1}c_m(x)y^m } 
\over 1+\sum_{m\geq1}m c_m(x) y^m } \big\}
             {1\over36} (\pdp F_0)  \cr
& \hskip2.5cm
+{1\over 36} (\pdp F_0) 
+{1\over36}\sum_{m\geq1} m {f_m \over \phi^2}y^{m-1} (\pdp y) \cr
&
={1\over 36}\(\pdp \log \big\{ (1-27x)+(1+y)^3-1\}^{-{1\over6}} 
   {{\rm e}^{-\sum_{m\geq1}c_m(x)y^m } 
 \over 1+\sum_{m\geq1}m c_m(x) y^m } \big\} \) \times (\pdp F_0) \cr
& \hskip2.5cm
+{1\over36} \pdp F_0+ {1\over 36} (\pdp)^2 F_0  \cr
&
={1\over 36} (\pdp F_1^{BCOV})(\pdp F_0) + {1\over36}\pdp(\pdp+1) F_0 \cr}
$$
Taking into account the difference of the normalization {\relationF}, we 
conclude the recursion relation. \hfill []

\vskip1cm

Now we may determine the generating function $Z_{1;n}(q)$ explicitly 
from {\FBCOV} under the relation $F_{1}(q,p)={1\over2} F_1^{BCOV}(q,p)$. 
As in the cese of genus zero, we may represent  $Z_{1;n}$ in terms of 
$f_1(x), \psi(x)$ and $E_2(q)$. Corresponding to {\Gnf}, we have
\eqn\GnfI{
Z_{1;n}=G_{1;n}(f_1,{E_2\over\phi^2})\phi^{2n} \( {f_1\psi \over \phi^2 } \)^n 
= G_{1;n}(f_1,{E_2\over\phi^2})\phi^{2n} \( Z_{0;1}\)^n \quad .
}
After straightforward evaluation of {\FBCOV}, we obtain for the first 
few of $G_{1;n}$'s;
$$
\eqalign{
&
G_{1;1}={1\over18}\( \phi^2 + {1\over2} E_2 \) \phi^{-2} \;\;,\;\;
G_{1;2}={1\over1728}
\( \Big(1+{24\over f_1}\Big)\phi^4 +{8\over3}E_2\phi^2 +{5\over3}E_2^2 \) 
\phi^{-4} \;\;, \cr
&
G_{1;3}={1\over15552} \( \Big(1+{48 \over f_1^2}\Big)\phi^6 + 
{13\over6}\Big(1-{ 8\over13 f_1} \Big)E_2 \phi^4 + E_2^2 \phi^2 
+ {13 \over 18}E_2^3 \) \phi^{-6} \;\;. \cr
}
$$
If we use the relations {\fEs} for the polynomials of $1/f_1$, we find 
$$
\eqalign{
&
G_{1;1}={1\over18}\(\phi^2 + {1\over2}E_2\)\phi^{-2} \;\;,\;\;
G_{1;2}={1\over432}\(\phi^4  - {1\over12}E_4 
         + {2\over3}\phi^2 E_2 + {5\over12}E_2^2 \)\phi^{-4}  \;, \cr
&
G_{1;3}={1\over7776}\( \phi^6 - {1\over6}\phi^2 E_4 - {1\over27}E_6 
         + E_2\Big(\phi^4 + {1\over108} E_4\Big) 
         + {1\over2}\phi^2 E_2^2 + {13\over36}E_2^3  \)\phi^{-6} \;\;. \cr
}
$$
Contrary to the case of $g=0$, the $\phi$-dependence remains in $Z_{1;n}(q)$ 
after the elimination of the polynomials of $1/f_1$. Thus we arrive at 
the follwing weaker statements about $Z_{1;n}(q)$;

\vskip0.5cm
\noindent
{\bf Proposition} \prop\modular 
{\it 
The generation function $Z_{1,n}(q)$ in {\GnfI} takes the form
$$
Z_{1;n}(q)=P_{2n}(\phi,E_2,E_4,E_6) \(Z_{0;1}(q)\)^n \;\;,
$$
where $P_{2n}$ is a 'quasi-modular form' of weight $2n$ for the 
modular subgroup $\Gamma(3)$. }

\vskip0.5cm
\noindent
{\bf Remark} \prop\RemI
\item{1)} The form of the polynomial $P_{2n}$ is not unique because of 
the relation {\fEs} among $\phi, E_4$ and $E_6$. \par\noindent
\item{2)} The function $Z_{1;n}$ contains the contribution from the 
genus zero curves (i.e., the degenerated instanton {\BCOVi}{\FP},) as well 
as the contribution from the multiple covers. We may separate these 
in $Z_{1;n}$ as follows;
\eqn\ZnsubI{
Z_{1;n}(q)=\tilde Z_{1;n}(q)+\sum_{k|n, k\not=1} 
\sigma_{-1}(k)\tilde Z_{1;n/k}(q^k)+
{1\over12}\sum_{k|n} {1\over k}\tilde Z_{0;n/k}(q^k) \;\;,
}
where $\sigma_{-1}(k)=\sum_{m|k} {1\over m}$. The function $\tilde Z_{1;n}(q)$ 
is expected to 'count' the numbers of the elliptic curves ${\cal C}$ with 
$({\cal C},F)=n$ in $S$. As is the case of $g=0$, we have certain vanishing 
conditions for the elliptic curves, which is useful to determine the 
integration constants for our holomorphic anomaly equation {\recZngI}. 
However as we will argue in the next section the appearance $\phi$ in the 
polynomial $P_{2n}$ increases unknown parameters in the integration 
constants. From this reason holomorphic anomaly equation become less 
powerful than the case of $g=0$ to determine $Z_{1;n}$.

\vfill\eject

\newsec{ Predictions for Gromow-Witten invariants of higher genera }
\global\global\propno=1

\subsec{ General considerations }

{}From the analysis for $g=0$ and $g=1$, we naturally come to the 
following conjecture about the holomorphic anomaly equation for all genera. 

\vskip0.5cm
\noindent
{\bf Conjecture} \prop\conjHol 
(Holomorphic anomaly equation for all $g$) 
{\it The generating function $Z_{g;n}(q)$ has the form 
\eqn\weakquasimod{
P_{2g+2n-2}(\phi,E_2,E_4,E_6) \(Z_{0;1}(q)\)^n 
}
with some 'quasi-modular form' for $\Gamma(3)$ of weight $2g+2n-2$. (In the 
special case of $g=0$, it simplifies to $P_{2n-2}(E_2,E_4,E_6)$, i.e., 
exactly the quasi-modular form of weight $2n-2$). And it  
satisfies the recursion relation
\eqn\holoano{
{\pd Z_{g;n} \over \pd E_2} = {1\over 72} \sum_{g'+g''=g} \sum_{s=1}^{n-1} 
s(n-s) Z_{g';s} Z_{g'';n-s} +{n(n+1) \over 72 } Z_{g-1;n} \;\;.
} }

\vskip0.3cm
In the following we will consider the solutions of the 
holomorphic anomaly equation. For this purpose, first of all, 
we need to have some data to fix the 'integration constants' for our recursion 
relation {\holoano}. Let us suppose that a curve ${\cal C}_g$ 
in the rational elliptic surface is in a homology class $[{\cal C}_g]
=d H - a_1 e_1 -a_2 e_2 \cdots - a_9 e_9$, where $e_i$'s refer to the 
$-1$ curves from the blowing ups. Then, since $Z_{g;n}$ counts 
the Gromov-Witten invariants of genus $g$ curves with 
$([{\cal C}_g], F)=n$, we should have 
\eqn\nsection{
([{\cal C}_g],F)=3d-\sum_{i=1}^9 a_i = n \;\;, } 
and for the arithmetic genus
\eqn\agenus{
g_a({\cal C}_g)={(d-1)(d-2) \over2}-\sum_{i=1}^9 {a_i(a_i-1)\over2} =g \;\;.}
If the curve ${\cal C}_g$ is singular, the arithmetic genus might be different 
from the genus of the normalization of ${\cal C}_g$. We will come to 
this point later, however for the moment we will ignore this difference. 
Then it is easy to see that the above two constrains provide 
us several vanishing conditions on the numbers of curves. 
In Table 1 we have presented the lowest degree $d=([{\cal C}_g],H)$ for 
which a curve ${\cal C}_g$ may exist for given $g$ and $n$. 
\vbox{
$$
\vbox{ \offinterlineskip \tabskip0.3cm
\halign{  \strut#&   \hfil#\hfil   &\vrule#  && \hfil # \hfil \cr 
 & $n \backslash g $ &&
           0 & 1 & 2 & 3 &  4 &  5 &  6 & 7 & 8 & 9 \cr
\noalign{\hrule}
 &   1  && 0 & 3 & 6 & 9 & 12 & 15 & 18 & 21& 24& 27   \cr
\noalign{\vskip-0.2cm}
 &   2  && 1 & 3 & 4 & 6 &  7 &  9 & 10 & 12& 13& 15 \cr
\noalign{\vskip-0.2cm}
 &   3  && 1 & 3 & 4 & 4 &  6 &  7 &  7 & 9 & 10& 10 \cr
\noalign{\vskip-0.2cm}
 &   4  && 2 & 3 & 4 & 4 &  5 &  6 &  7 & 7 & 8 & 9  \cr
\noalign{\vskip-0.2cm}
 &   5  && 2 & 3 & 4 & 4 &  5 &  5 &  6 & 7 & 7 & 8  \cr
}}
$$
{\leftskip1cm\rightskip1cm 
{\bf Table 1}. Each number shows the lowest degree $d=([{\cal C}_g],H)$ 
for curves of given $g$ and $n$. \par}
}
\vskip1cm

The homology classes of curves in the table may be written explicitly  
for given $g$ and $n$. For example, for $n=1$ they are simply given by
$[{\cal C}_g]= e_i + g F \;\;\;\; (i=1,\cdots,9) \;. $
The data in the table provides us vanishing conditions for the 
first few terms in the $q$-expansion of $\tilde Z_{g;n}(q)$, where tilde 
represents the subtraction of the {\it degenerated instantons} 
from the Gromov-Witten invariants. We may relate $Z_{g;n}(q)$ $(g\geq2)$ 
to the subtracted functions by
\eqn\subtractZ{
Z_{g;n}(q)=\tilde Z_{g;n}(q) + 
\sum_{k|n,k\not=1} D(g,g,k)\tilde Z_{g;n/k}(q^k) +
\sum_{h=0}^{g-1} \sum_{k|n}D(g,h,k) \tilde Z_{h;n/k}(q^k) \;\;, }
with some rational numbers $D(g,h,k)$. Therefore all we need to fix from the 
vanishing conditions are the 'integration constants' together with the
rational numbers $D(g,h,k)$. 
For higher $n>4$ it turns out that 
the vanishing conditions in Table 1 are not sufficient to determine 
completely both the integration constants of the recursion {\holoano} 
and the form of the degenerated instanton $D(g,h,k)$. However we will 
see based on a simple counting arguments that for lower $n\leq 3$ 
they suffices at least to fix the integration constants for our 
recursion relation. Especially for the case $n=1$ they determine 
both the integration constants and the form of the degenerated instanton. 

As an extreme case let us first consider the genus zero 
generating functions $Z_{0;n}(q) \; (n=1,2,\cdots)$.  
As we have already considered in Remark {\RemMulti},  
the only undetermined in this case are the integration constants 
in the polynomial $P_{2n-2}$ in {\ZnP}. 
We see that these constants can be fixed by the vanishing conditions 
in Table 1 comparing the first column of the table with the 
dimensionality of the modular form of a given weight, 
which is given by the series
\eqn\growthEE{
{1\over (1-t^4)(1-t^6)}=  
1 + {t^4} + {t^6} + {t^8} + {t^{10}} + 2\,{t^{12}} + {t^{14}} + 
   2\,{t^{16}} + 2\,{t^{18}} + 2\,{t^{20}}  +\cdots \;\;.
}
Now let us look at the functions $Z_{g;n}(q)$ for all $g$ and 
lower $n (\leq 3)$. 
In this case we only have the weaker assumption on  
$P_{2g+2n-2}$ in {\weakquasimod}. Taking into account the relation among 
$\phi, E_4$ and $E_6$ in the second line of {\fEs}, 
we may estimate the relevant dimensionality of the 
modular form $P_{2g+2n-2}|_{E_2=0}$ (, integration constants of the 
holomorphic anomaly equation,) by
\eqn\growthPEE{
{(1+t^2+t^4+t^6)\over (1-t^4)(1-t^6)} = 
 1 + {t^2} + 2\,{t^4} + 3\,{t^6} + 3\,{t^8} + 4\,{t^{10}} + 5\,{t^{12}} + 
   5\,{t^{14}} + 6\,{t^{16}} + 7\,{t^{18}} + 7\,{t^{20}} + \cdots \;\;.
}
The growth of the dimensions should be compared with each line of 
the table 1 under a suitable shift. From these comparisons of the 
numbers of the 'integration constants' and the numbers of the vanishing 
conditions, we may deduce that for $n\leq 3$ the vanishing conditions 
suffices to determine the integration constants while leaving some  
of $D(g,h,k)$ undetermined. To go beyond this rather unsatisfactory situation, 
we need to know more details about the numbers of genus $g$ curves of 
a given homology classes or the form of the degenerated instanton 
$D(g,h,k)$. Though our simple vanishing conditions are insufficient 
to determine all the unknowns, we see from the first line of the table 
they are restrictive enough to fix the form of $Z_{g;1}(q)$ and 
$D(g,h,1)$ completely. In the next subsection we will present a detailed 
analysis of $Z_{g;1}(q)$ for all $g$.

\subsec{ Gromov-Witten invariants $Z_{g;1}$ and degenerated instantons}

To get some intuition about the curves ${\cal C}_g$ in $S$, let us 
first recall the form of $Z_{0;1}(q)$ obtained in {\HSS}{\Zagier} 
\eqn\hssZO{
Z_{0;1}(q)={q^{3\over2}\Theta_{E_8}(3t, t\gamma) \over \eta(q^3)^{12} }
 \;\;\; \( = 9 {q^{1\over6} \over \eta(q)^4 } \)
}
where $\gamma=(1,\cdots,1,-1)$. The appearance of the $E_8$ theta function 
originates from the well-established fact that the sections of the 
rational elliptic surface form additive group called Mordell-Weil 
group, and it becomes a lattice isomorphic to the $E_8$ lattice 
endowed with a positive definite bilinear form{\Shi}{\HSS}{\SaII}. 
The eta functions in the denominator have been explained by introducing 
{\it pseudo-section} which is a composite of a section with 
some of the twelve singular fibers with its homology class $\sigma + 
k F$. Thus the function $Z_{0;1}$ counts the numbers of pseudo-sections 
in $S$ which are not irreducible but naturally come in the 
theory of the stable maps{\BL}. 

Now in our general case of genus $g$ curves with $({\cal C}_g, F)=1$ the 
function $Z_{g;1}$ counts the genus $g$ sections of the elliptic fibration 
of $S$. Since the generic fiber spaces are elliptic curves, the genus $g$ 
section are composite of two components, one is a pseudo-section and 
the other consists of $g$ fiber elliptic curves. The genus $g$ sections 
of the lowest degree are those with their homology classes given by
\eqn\sectionCg{
[{\cal C}_g]= e_i + g F \;\;\;\; (i=1,\cdots,9) \;. }
Thus the expansion of $Z_{g;1}(q)$ start from $q^{3g}$ with its 
coefficient 'counting' the number of the genus $g$ 
sections of class {\sectionCg}. (This is the vanishing condition 
we have listed in Table 1 for $n=1$ and $g$.) 
The $g$ elliptic curves in general avoid the twelve singular fibers and 
make a $g$-dimensional family parameterized by ${\rm Sym}^g(\IP^1)$. 

As we have already remarked, the vanishing conditions for 
the 'numbers' of curves grow much faster than the dimensionality of 
the integration constants {\growthPEE} plus the numbers of the unknowns 
$D(g,h,1)$ in 
\eqn\subtractZnI{
Z_{g;1}(q)=\tilde Z_{g;1}(q) + 
\sum_{h=0}^{g-1} D(g,h,1) \tilde Z_{h,1}(q) \;\;. }
Owing to this nice property, we can integrate our holomorphic anomaly 
equation for $n=1$, 
\eqn\holoanoI{
{\pd Z_{g;1} \over \pd E_2} = {1\over 36} Z_{g-1;1}  \;\;, }
with the results listed in Table 2 for $\tilde Z_{g;1}(q)$'s. 
For the degenerated instantons $D(g,h,1)$ we find
\eqn\degD{
\eqalign{
 & Z_{1;1}=\tilde Z_{1;1}+{1\over12}\tilde Z_{0;1} \cr
 & Z_{2;1}=\tilde Z_{2;1} + \chi(M_2) \tilde Z_{0;1} \cr
 & Z_{3;1}=\tilde Z_{3;1} + {1\over3!}\chi(M_3)\tilde Z_{0;1}
        -{1\over12}\tilde Z_{1;1}  \cr
 & Z_{4;1}=\tilde Z_{4;1} + {1\over5!}\chi(M_4)\tilde Z_{0;1}
        +{1\over360}\tilde Z_{2;1}-{1\over6}\tilde Z_{3;1} \cr
 & Z_{5;1}=\tilde Z_{5;1} + {1\over7!}\chi(M_5)\tilde Z_{0;1}
        -{1\over20160}\tilde Z_{2;1}+{1\over80}\tilde Z_{3;1}
        -{1\over4}\tilde Z_{4,1} \;\;,\cr}  } 
with the orbifold Euler number of the moduli space of genus $g$ stable 
curves, $\chi(M_g)={ |B_{2g}| \over 2g(2g-2)!} (g\geq 2)$. 
In fact these forms of 
the degenerated instantons $D(g,h,1) \; (0\leq h < g)$  coincide with 
recently established results in {\GVii}{\Pa}, where we have
\eqn\GVdeg{
Z_{g;1}(q)=\sum_{h=0}^g C_h(g-h,1) \tilde Z_{h;1}(q) \;\;,
}
with the coefficients determined by 
\eqn\Cgh{
\( {\sin(t/2)\over t/2} \)^{2g-2} 
= \sum_{h=0}^\infty C_g(h,1) t^{2h} \;\;\;.}
 
\vskip0.5cm

$$
\vbox{ \offinterlineskip 
\halign{  \strut \vrule# & $\;$  #   \hfill &\vrule# \cr 
\noalign{\hrule}
& 
$\tilde Z_{0;1}$=$ \quad
9+36q+126q^2+360q^3+945q^4+2268q^5+5166q^6+11160q^7+\cdots $
&\cr
\noalign{\hrule}
&
$ \tilde Z_{1;1}$=$
-18q^3-72q^4-252q^5-774q^6-2106q^7-5292q^8-12564q^9-\cdots $
&\cr
\noalign{\hrule}
&
$ \tilde Z_{2;1}$=$\;\;\;
27q^6+108q^7+378q^8+1224q^9+3411q^{10}+8820q^{11}
     +21663q^{12}+\cdots $
&\cr
\noalign{\hrule}
&
$ \tilde Z_{3;1}$=$
-36q^9-144q^{10}-504q^{11}-1710q^{12}-4860q^{13}-12852q^{14}-\cdots $
&\cr
\noalign{\hrule}
&
$ \tilde Z_{4;1}$=$\;\;\;
45q^{12}+180q^{13}+630q^{14}+2232q^{15}+6453q^{16}+17388q^{17}+\cdots $
&\cr
\noalign{\hrule}
&
$ \tilde Z_{5;1}$=$
-54q^{15}-216q^{16}-756q^{17}-2790q^{18}-8190q^{19}-22428q^{20}- \cdots $
&\cr
\noalign{\hrule} } 
}
$$
{\leftskip1cm\rightskip1cm
{\bf Table 2}. Solutions of the holomorphic anomaly equation {\holoanoI}. 
These are related to $Z_{g;1}$ by {\degD}. 
\par}

\vskip1cm

We note in Table 2 that for the first three terms in the expansion 
we have 
\eqn\expandZ{
\tilde Z_{g;1}(q) = (-1)^g \chi({\rm Sym}^g(\IP^1))( 9 q^g+36q^{g+1}
 +126 q^{g+2} + \cdots  )
}
where $\chi({\rm Sym}^g(\IP^1))$ represents the Euler number of 
${\rm Sym}^g(\IP^1)=\IP^g$. This is also in agreement with the argument 
in {\GVii} for counting curves with moduli. From the fourth term 
in the series expansion {\expandZ}  
contributions from the singular fibers come in, which somehow 
generalize the situation we encountered in the case of $g=0$. 

For the first few of $Z_{g;1}(q)$'s we have determined explicitly the forms 
of the polynomials $P_{2g}(\phi,E_2,E_4,E_6)$. We find that if we use the 
following relations 
\eqn\phiE{
\eqalign{
&3E_2(q^3)=2\phi(q)^2+E_2(q)\;\;,\;\; 
 9E_4(q^3)=10 \phi(q)^4 -  E_4(q) \;\;,\;\; \cr
& \quad 27 E_6(q^3)=35 \phi(q)^6-7 E_4(q)\phi(q)^2-E_6(q) \;\;,\cr }}
they summarize into concise forms;
\eqn\ZEphi{
\eqalign{
& Z_{0;1}={q^{3\over2}\Theta_{E_8}(3t, t\gamma) \over \eta(q^3)^{12} }
  \;\;,\qquad  
  Z_{1;1}={1\over12} E_2(q^3) Z_{0;1}  \cr
&  Z_{2;1}={1\over1440}\big( 5\,E_2(q^3)^2+E_4(q^3) \big)Z_{0;1} \cr
& Z_{3;1}={1\over362880} \big(35 E_2(q^3)^3+21 E_2(q^3)E_4(q^3)+4E_6(q^3)
            \big) Z_{0;1}    \;\;\;.
\cr} }

\noindent
{\bf Proposition} \prop\conjP 
{\it The solutions of the holomorphic anomaly equation {\holoanoI} take 
the following general form
$$
Z_{g;1}(q)=P_{2g}(E_2(q^3),E_4(q^3),E_6(q^3)) \; Z_{0;1}(q) \;\;,
$$
where $P_{2g}$ is a quasi-modular form of weight $2g$. }

\vskip0.3cm
The reason of this simplification will be explained in the next section. 
Here for later use we define ${\cal G}_{g;1}$ by
\eqn\calG{
\tilde Z_{g;1}(q) = \Theta_{E_8}(3t,t\gamma) \; {\cal G}_{g;1}(q^3)  \;\; ,}
which should count the genus $g$ pseudo-sections made from a section, say 
the zero section. We verify directly that the functions ${\cal G}_{g;1}$'s 
depend on $q$ through $q^3$ and have the following expansions;
\eqn\calGexp{
\eqalign{
&{\cal G}_{0;1}=  \;\;\;
   1 + 12\,{q^3} + 90\,{q^6} + 520\,{q^9} + 2535\,{q^{12}} + 10908\,{q^{15}} + 
   42614\,{q^{18}} + \cdots \cr 
&{\cal G}_{1;1}=  
{} -2\,{q^3} - 30\,{q^6} - 260\,{q^9} - 1690\,{q^{12}} - 9090\,{q^{15}} - 
   42614\,{q^{18}} - \cdots  \cr
&{\cal G}_{2;1}= \;\;
   3\,{q^6} + 52\,{q^9} + 507\,{q^{12}} + 3636\,{q^{15}} + 21307\,{q^{18}} + 
   107772\,{q^{21}} + \cdots \cr 
&{\cal G}_{3;1}= 
{} -4\,{q^9} - 78\,{q^{12}} - 840\,{q^{15}} - 6570\,{q^{18}} - 
   41580\,{q^{21}} - 225432\,{q^{24}} - \cdots \cr
&{\cal G}_{4;1}= \;\;
   5\,{q^{12}} + 108\,{q^{15}} + 1271\,{q^{18}} + 10756\,{q^{21}} + 
   73083\,{q^{24}} + \cdots \cr
&{\cal G}_{5;1}= 
{} -6\,{q^{15}} - 142\,{q^{18}} - 1812\,{q^{21}} - 16494\,{q^{24}} - 
   119770\,{q^{27}} - \cdots \cr
} } 
In the next section we will discuss geometric interpretation of 
the numbers in the above expansions.

\vfill\eject

\newsec{ Discussions }
\global\global\propno=1

\subsec{ Counting BPS states }

In this section we would like to discuss relations of our results 
to the very interesting proposals made in the recent works 
by Gopakumar and Vafa {\GVi}{\GVii}.

It is known in physics that the genus $g$ prepotential $F_g$ has 
a meaning as the genus $g$ topological partition function of the 
twisted Calabi-Yau sigma model in the type IIA string theory and it 
does not receive the string perturbative corrections due to the 
fact that the dilaton belongs to the hypermultiplet in the type IIA string.   
Since the heterotic/type II string duality connects the heterotic dilaton 
field to one of the vectormultiplet moduli in the type IIA side, 
we may expect to extract the non-perturbative properties in 
the type IIA side from the perturbation theory in the heterotic string. 
In our case of the topological amplitude $F_g$, Gopakumar and Vafa 
{\GVi}{\GVii} have found that it can be derived from the one loop 
integral in the heterotic side. They found that the Schwinger one-loop 
calculation for a particle in a constant background electro-magnetic 
filed applied to the BPS states with spin $(j_1,j_2)$ under 
$SO(4)=SU(2)_L\times SU(2)_R$ determines the higher genus $F_g$. 
According to {\GVii} the contribution of each BPS state with spin $(j_1,j_2)$ 
to $F_g$ is determined by the following decomposition with respect to 
$SU(2)_L \subset SU(2)_L\times SU(2)_R$;
\eqn\GVBPSjj{
(j_1,j_2)= 
\sum_{r=0}^{2j_1} \alpha_r [({1\over2})\oplus 2(0)]^{\otimes r}  \;\;,
}
where we allow $\alpha_r$ formally to be {\it negative} integer. Then the 
topological partition function has the following expression 
\eqn\GVBPS{
\sum_{g=0}^\infty \lambda^{2g-2} F_g 
=\sum_{\Gamma: BPS} \; \sum_{r=0}^{2j_1(\Gamma)} \; \sum_{m\in \ZZ} 
\alpha_r(\Gamma)  
\int_\epsilon^\infty {ds \over s} \(2 \sin{s\over2}\)^{2r-2} 
{\rm e}^{-2\pi {s\over\lambda}(A(\Gamma)+i m)} \;\;, 
}
where $A$ measures the central charge of the BPS state $\Gamma$ and 
the summation over $m$ is explained as the central charge of 
the fifth dimension which originates in the M-theory. In the type IIA theory 
the central charge $A$ is measured 
by the K\"ahler classes of the corresponding Calabi-Yau manifolds, 
and after the integration we arrive at
\eqn\GVFall{
\sum_{g=0}^\infty \lambda^{2g-2} F_g(t) = 
\sum_{0\not=\eta \in H_2(X,\ZZ)} \; \sum_{r \geq 0} \; \sum_{k>0} 
\alpha^\eta_r {1\over k} \(2 \sin{k \lambda \over2}\)^{2r-2} q^{k\eta} \;\;,
}
where $\alpha^\eta_r$ represents the summation of $\alpha_r(\Gamma)$ over 
the BPS states $\Gamma$ with charge $\eta$,  and 
$q^{k\eta}={\rm exp}(-2\pi (k \eta, K))$ with 
$K=t_1J_1+\cdots+t_{h^{1,1}} J_{h^{1,1}}$ the K\"ahler class. 
Expanding {\GVFall} with respect to $\lambda$ we obtain
\eqn\GVFg{
F_g(q)= \sum_{0\not=\eta \in H_2(X,\ZZ)} \; \sum_{h=0}^g 
\; \sum_{k | \eta} \alpha_h^{\eta/k} k^{2g-3} C_h(g-h,1) q^\eta  \;\;,
}
which is the general form proposed in {\GVii} and we have reproduced 
in our special case (, see {\degD} and {\Cgh}). 

In the type IIA picture the BPS state appears as the D2 brane with 
the flat $U(1)$ connection. Let us suppose that a genus $g$ curve in 
a Calabi-Yau manifold comes with the moduli of the deformation ${\cal M}$. 
Together with the Jacobian of each curve we have the fiber space 
$\hat {\cal M} \rightarrow {\cal M}$ for the BPS states. 
Gopakumar and Vafa argued based on the M-theory that in this case 
the counting BPS states in {\GVBPS} is interpreted by 
the Lefschetz $SL(2,\IC)$ decomposition of the 
cohomology of the moduli space $\hat {\cal M}$. They propose 
the Lefschetz decomposition with respect to the fiber $SL(2,\IC)_L$ and 
the base $SL(2,\IC)_R$ of the cohomology $H^{*}(\hat {\cal M})$. Once we 
assume existence of these  $SL(2,\IC)$ actions\footnote{$^\dagger$}
{One of the Lefschetz $SL(2,\IC)$ actions is the multiplication of the 
K\"ahler form $k$. Since our moduli spaces have a natural fibration structure 
$\pi: \hat {\cal M}(g) \rightarrow {\cal M}(g)$ with a section $\iota$, 
we may decompose the K\"ahler class $k$ into $k_L=(k-\pi^*(\iota^*(k)))$ 
and $k_R= \iota^*(k)$. This decomposition defines the $SL(2,\IC)_L\times 
SL(2,\IC)_R$ actions. We would like to thank Y. Shimizu for pointing 
this out to us.} 
, we may arrange this 
decomposition as
\eqn\lefschetz{
I_g\otimes R_g + I_{g-1}\otimes R_{g-1} + \cdots + I_0\otimes R_0 \;\;,}
where $I_k:=[({1\over2})\oplus 2(0) ]^{\otimes k}$. Identifying this 
decomposition with that in {\GVBPSjj}, they propose 
\eqn\aRk{
\alpha_k = \chi(R_k) \quad (k=0,\cdots,g) \;,
}
where $\chi(R_k)$ is the dimensions of the $SL(2,\IC)_R$ representations 
in $R_k$ weighted with $(-1)^{2j_R}$. As argued in {\GVii} we can determine 
$\alpha_g$ and $\alpha_0$ easily by the geometric Euler numbers;
\eqn\algo{
\alpha_g=(-1)^d \chi({\cal M}) \;\;,\;\; 
\alpha_0=(-1)^{\hat d} \chi(\hat {\cal M}) \;\;,}
where $d={\rm dim}{\cal M}$ and $\hat d={\rm dim}\hat{\cal M}$. 

Now let us consider our genus $g$ pseudo-sections with a fixed section, say 
the zero section. Counting BPS states for these pseudo-sections 
are summarized in the generating function ${\cal G}_{g;1}$ in {\calGexp}. 
As we have seen in the last section the genus $g$ pseudo-sections 
come with the moduli ${\cal M}(g):={\rm Sym}^g(\IP^1)=\IP^g$. 
Furthermore the data of the Jacobian may be specified  
by $g$ points on $S$, one for each elliptic fiber. Therefore we 
naturally come to the space 
\eqn\symS{
\hat{\cal M}(g)=\widetilde{\rm Sym}^g(S) \;\;, }
where $\tilde{\;}$ represents the resolution of the orbifold singularities 
via the Hilbert scheme of $g$ points on $S$, which we denote $S^{[g]}$. The 
construction of the Lefschetz actions on this space and making the 
decomposition {\lefschetz} would be interesting problem, however we 
already have predictions for the decomposition; 
\eqn\lefR{
I_g \times \chi(R_g)+I_{g-1} \times \chi(R_{g-1}) + \cdots 
+ I_0 \times \chi(R_0) \;\;.}  
Namely we can read {\lefR} in our expansion {\calGexp}:  
\eqn\predict{
\eqalign{
g=1 & \hskip1cm -2 \, I_1 + 12 \, I_0 \cr
g=2 & \hskip1.4cm  3 \, I_2-30 \,I_1 + 90 \,I_0 \cr
g=3 & \hskip1cm  -4 \,I_3 + 52 \,I_2 -260 \,I_1 + 520 \,I_0  \cr
g=4 & \hskip1.4cm  5 \,I_4 -78 \,I_3 + 507 \,I_2 - 1690 \,I_1 + 2535 \,I_0  \cr
g=5 & \hskip1cm  -6 \,I_5 + 108 \,I_4 - 840 \,I_3 
                    + 3636 \,I_2 - 9090 \,I_1 + 10908 \,I_0  \cr }}
In the next subsection we propose a natural generalization of 
G\"ottsche's formula for the Poincar\'e polynomials of $S^{[g]}$ and 
reproduce the above predictions. Our generalization of G\"ottsche's 
formula suffices to determine the decomposition {\lefschetz}.

\vskip1cm

\subsec{ G\"ottsche's formula with $SL(2,\IC)_L\times SL(2,\IC)_R$ }

G\"ottsche's formula describes the generating function for the 
Poincar\'e polynomials of the Hilbert scheme of $g$ points on 
a surface $S$. In our case it appears as a natural resolution $S^{[n]}$  
of the symmetric product ${\rm Sym}^g(S)$ for the rational elliptic 
surface. If we assume the existence of the Lefschetz actions 
$SL(2,\IC)_L\times SL(2,\IC)_R$, then the Poincar\'e polynomial 
can be written 
\eqn\PSn{
P_t(S^{[g]})=(t^{2g})\; {\rm Tr}_{H^*(S^{[g]})} t^{2(j_{3,L}+j_{3,R})} 
}
in terms of the diagonal $SL(2,\IC)$ action. Then the problem is to  
recover both left and right charges in the above formula, namely, 
$P_{t_L,t_R}(S^{[g]})= (t_L^{g}t_R^{g})
{\rm Tr} t_L^{2j_{3,L}}t_R^{2j_{3,R}}$. In the case of $g=1$, 
we note that the decomposition is unique as follows
\eqn\LRdecomp{
\big({1\over2},{1\over2}\big)_{L,R} \oplus 8 \big(0,0\big)_{L,R} 
=
\big(1\big)_{L+R} \oplus 9 \big(0\big)_{L+R} 
}

Now let us recall G\"ottsche's formula{\G} 
\eqn\Goettsche{
G(t,q)=\prod_{n\geq1} {1 \over 
\(1-t^{2n-2}q^n\)\(1-t^{2n}q^n\)^{10}\(1-t^{2n+2}q^n\) } \;\;.
}
As has been interpreted in {\VW} there is a close relation between 
(co)homology elements and the bosonic oscillators associated to each 
elements in $H^*(S)$. Under this correspondence the classical cohomology 
is represented by the lowest modes, say $a_k(-1) \; (k=1,\cdots,12)$, 
and generate the symmetric product of $H^*(S)$. The higher mode excitations 
$a_k(-m)$ come from the singular strata of the point configurations. 
Here it is natural to assume that the higher mode excitation $a_k(-m)$ 
have the same spin as the lowest mode $a_k(-1)$, whose spin contents 
are uniquely determined in {\LRdecomp}. Under this assumption it is 
easy to recover the  $SL(2,\IC)_L \times SL(2,\IC)_R$ spin weights 
in G\"ottsche's formula
\eqn\GoettscheLR{
\eqalign{
G(t_L,t_R,q)=&\prod_{n\geq1} \bigg\{ {1 \over 
\(1-(t_Lt_R)^{n-1}q^n\)
\(1-(t_Lt_R)^{n+1}q^n\)} \cr
& \times { 1\over 
\(1-t_L^2(t_Lt_R)^{n-1}q^n\)\(1-t_R^2(t_Lt_R)^{n-1}q^n\)
\(1-(t_Lt_R)^{n}q^n\)^{8} } \bigg\} \;\;, \cr}
}
which provides us the Poincar\'e polynomial $P_{t_L,t_R}(S^{[g]})$ 
as the coefficient of $q^g$.  Then our predictions (5.10) 
for the Lefschetz decomposition should be verified in the formula 
\eqn\index{
{1\over (t_L t_R)^g} P_{t_L,t_R}(S^{[g]}) \bigg|_{t_R=-1} 
=\chi(R_g) \(t_L+{1\over t_L}+2\)^{g} 
+ \chi(R_{g-1})\(t_L+{1\over t_L}+2\)^{g-1}
+\cdots+ \chi(R_{0}) \;\;.
}
For lower $g$, we have found complete agreements of our predictions 
from the generating functions ${\cal G}_{g;1}$ with those coming 
from the above formula. For example, for $g=3$ we obtain from {\index}
$$
{}-4 \(t_L+{1\over t_L}+2\)^3 + 52 \(t_L+{1\over t_L}+2\)^2 
{}-260 \(t_L+{1\over t_L}+2\) +520\;,
$$
which should be compared with {\predict}. We may summarize the above results 
into a formula relating G\"ottsche's formula with our generating 
functions{\calG};
\eqn\GcalG{
G(t_L,-1,-{q^3\over t_L})=\sum_{g\geq 0} {\cal G}_{g;1}(q^3) 
\(t_L+{1\over t_L}+2\)^{g} \;\;.
}

\vskip1cm
\subsec{ Topological string partition function}

So far we have fixed a section to discuss the moduli space of 
the curves ${\cal C}_g$. To recover the contributions from the 
Mordell-Weil lattice, we simply need to multiply the $E_8$ theta function 
to the functions we have discussed. Therefore for the BPS state counting 
on the rational elliptic surface $S$ we consider the function 
$ \Theta_{E_8}(3t,t\gamma) G(-t_L,-1,{q^3 \over t_L})$ in terms of the 
generalized G\"ottsche's formula. Now it is easy to deduce the following 
relation;
\eqn\partition{
\eqalign{
\Theta_{E_8}(3t,t\gamma) G(-t_L,-1,{q^3 \over t_L})
&= \sum_{g\geq 0} \tilde Z_{g;1}(q) (-t_L-{1\over t_L} +2)^g \cr
&= \sum_{g\geq 0} Z_{g;1}(q) \lambda^{2g-2} \; 
(2{\rm sin}{\lambda\over2})^2 \;\;,\cr
} }
where $t_L={\rm e}^{i \lambda}$ and we have used the form of the degenerated 
instantons given by {\GVdeg}. 
This implies the function $ \Theta_{E_8}(3t,t\gamma) 
G(-t_L,-1,{q^3 \over t_L})$ 
provides us the all genus topological partition function. We may write 
{\partition} explicitly by
\eqn\topologicalZ{
q^{3\over2} {\Theta_{E_8}(3t, t\gamma) \over \eta(q^3)^{12} }
\prod_{n\geq1} {(1-q^{3n})^4 \over (1-t_L q^{3n})^2 (1-{1\over t_L} q^{3n})^2}
= \sum_{g\geq 0} 
Z_{g;1}(q) \lambda^{2g-2} (2{\rm sin} {\lambda\over2})^2\;\;. }  
Here we recognize the helicity generating function in the left hand 
side. Thus the topological partition function has a simple but suggestive form 
that the genus zero function $Z_{0;1}$ multiplied by the helicity generating 
function, which is actually the starting point of the analysis done in 
{\GVii}. 

Here for completeness we prove the compatibility of the above 
result {\partition} with our 
holomorphic anomaly equation {\holoanoI}. For this purpose we note the 
following identity which can be proved in a straightforward way;
\eqn\helE{
\({ {\lambda/2} \over {\rm sin}\lambda/2 } \)^2 
\prod_{n\geq1} { (1-q^n)^4 \over
(1-{\rm e}^{i \lambda} q^n)^2 (1-{\rm e}^{-i \lambda} q^n)^2 }
= 
{\rm exp}\(2 \sum_{k\geq 1} {\zeta(2k) \over k} E_{2k}(q) 
\({\lambda \over 2\pi }\)^{2k} \) \;\;. }
The compatibility may be easily verified if we use 
$3E_2(q^3)=2\phi(q)^2+E_2(q)$ found in {\phiE} and the value 
$\zeta(2)={\pi^2 \over 6}$, 
since we have ${\lambda^2 \over 36}$ for the both side of {\topologicalZ} 
after the differentiation with respect to $E_2(q)$. Also the formula {\helE} 
explains the simplification we have encountered in Proposition {\conjP}. 
Namely we have 
\eqn\ZgI{
\sum_{g\geq0}Z_{g;1}(q) \lambda^{2g}= 
Z_{0;1}(q) \; 
{\rm exp}\(2 \sum_{k\geq 1} {\zeta(2k) \over k} E_{2k}(q^3) 
\({\lambda \over 2\pi }\)^{2k} \) \;\;. }

\vfill\eject

\appendix{}{Picard-Fuchs equations of the mirror $X^\vee$}

Following {\HKTY}{\HLY} the Picard-Fuchs differential operators 
about the large complex structure limit are determined to be 
$$
\eqalign{
{\cal D}_1&
=9\tx^2-3\tx\ty-6\tx\tz+24\tz\ty-16\tz^2
 -27x(3\tx+\ty+2\tz+2)(3\tx+\ty+2\tz+1) \cr
& \hskip1cm -3y(\tx-8\tz)(\ty-\tz)+z(60\tx+32\tz+32)(3\tx+\ty+2\tz+1) \cr 
{\cal D}_2&
=\ty^2+y(3\tx+\ty+2\tz+1)(\ty-\tz) \cr
{\cal D}_3&
=(\tz-\ty)\tz-z(3\tx+\ty+2\tz+2)(3\tx+\ty+2\tz+1) \cr}
$$
where $\tx=x {\pd \over \pd x}$, etc.  Looking at the characteristic 
variety of this system we have determined the discriminant {\discrim}.

\vskip2cm
\centerline{\bf References }

\item{\BCOVi} \refBCOVi
\item{\BCOVii} \refBCOVii
\item{\Bat} \refBat
\item{\BM} \refBM
\item{\BL} \refBL
\item{\CdGP} \refCdGP
\item{\CN} \refCN
\item{\FP} \refFP
\item{\G} \refG
\item{\GVi} \refGVi
\item{\GVii} \refGVii
\item{\HKTY} \refHKTY
\item{\HLY} \refHLY
\item{\HSS} \refHSS
\item{\KMV} \refKMV
\item{\KM} \refKM
\item{\Ma} \refMa
\item{\MM} \refMM
\item{\MNW} \refMNW
\item{\MNVW} \refMNVW
\item{\MorI} \refMorI
\item{\MorII} \refMorII
\item{\Pa} \refPa
\item{\SaII} \refSaII
\item{\Shi} \refShi
\item{\VW} \refVW
\item{\YZ} \refYZ
\item{\Zagier} \refZagier


\bye